 \documentclass{aa} 
\usepackage{graphicx}
\usepackage{txfonts}
\begin{document}
   \title{Dust-to-gas ratios in the Kepler supernova remnant}

   \author{M. Contini
          \inst{1,2}
          }

   \offprints{M. Contini}

   \institute{Instituto de Astronomia, Geof\'{\i}sica e Ci\^encias
         Atmosf\'ericas - USP, Brazil
      \and
   School of Physics and Astronomy, Tel Aviv University, Tel Aviv
              69978, Israel 
             }

   \date{Received ; }

   \abstract{

A new method to evaluate the dust-to-gas ratios in the Kepler SNR
is presented. Dust emission  in the  infrared and bremsstrahlung
are calculated consistently, considering    that    dust grains are collisionally
heated by  the gas throughout the  front and downstream of both the expanding and the reverse shocks.
The  calculated continuum SED is  constrained by the observational data.  
The dust-to-gas ratios are determined by the ratio of the dust emission bump and bremsstrahlung in the  infrared.
The shell-like morphological similarity  of  X-ray    and radio emission, and of    the \Ha ~and  infrared images 
confirms that both radio  and X-ray emissions are created at the front of the  expanding  shock 
  and  that dust and gas are coupled   crossing  the expanding and reverse shock fronts.
The results show that large  grains with radius of $\sim$ 1 \mum ~with dust-to-gas ratios  $<$ 4 10$^{-3}$ 
survive sputtering and are  heated to a maximum temperature of 125 K 
 downstream of the shock expanding outwards with  velocity of about 1000 \kms.
The high velocity shocks become radiative for dust-to-gas ratios $>$ 10$^{-3}$.
Such shocks do not appear in the NE region, indicating that dust grains
are not homogeneously distributed throughout the remnant.
Smaller grains with radius of about  0.2\mum ~and   dust-to-gas ratios
of  $\sim$ 4 10$^{-4}$ are heated to a maximum temperature of $\sim$ 50 K
 downstream of the reverse shock corresponding to velocities of about 50 \kms.
A maximum dust mass   $<$  0.16 \msol is calculated. 


   }

%

\def\ea{\it et al. \rm}
\def\am{$^{\prime}$\ }
\def\as{$^{\prime\prime}$\ }
\def\msol{M$_{\odot}$ }
\def\kms{$\rm km\, s^{-1}$}
\def\cm3{$\rm cm^{-3}$}
\def\Ts{$\rm T_{*}$}
\def\Vs{$\rm V_{s}$}
\def\n0{$\rm n_{0}$}
\def\B0{$\rm B_{0}$}
\def\ne{$\rm n_{e}$}
\def\Te{$\rm T_{e}$}
\def\Tgr{$\rm T_{gr}$}
\def\Tgas{$\rm T_{gas}$}
\def\Ec{$\rm E_{c}$}
\def\Fh{$\rm F_{H}$}
\def\Hb{H$\beta$}
\def\Ha{H$\alpha$}
\def\Fn{$\rm F_{n}$}
\def\Fh{$\rm F_{h}$}
\def\erg{$\rm erg\, cm^{-2}\, s^{-1}$}
\def\mum{$\mu$m}
\def\Lx{L$_X$~}
\def\Fx{F$_X$}
\def\LIR{L$_{IR}$~}
\def\L12{L$_{12\mu m}$~}
\def\F12{F$_{12\mu m}$~}
\def\agr{a$_{gr}$}
\def\mm{$\mu$m}

   \maketitle
---------------------------------------------------------------------

\section{Introduction}

Since   the first identification by Baade (1943)
the remnant of the Kepler  supernova (SN)  which exploded in 1604, 
has  been  observed in the different wavelength ranges.

 van den Bergh \& Kemper (1977) studied
brightness variations and proper motions of the filaments in the supernova remnant (SNR).
Spectrophotometry data of the bright-western knots 
in the spectral range 3700-10,500 \AA ~were  presented  by Dennefeld (1982).
Observations in the optical range for single bright knots
were obtained by Leibowitz \& Danziger (1983) and by Blair, Long \& Vancura (1991). 

Data in the X-ray  (e.g. Hughes 1999) and in the radio (e.g. DeLaney et al. 2002) ranges
were also recently presented. 
Morgan et al (2003) provided  SCUBA data of the continuum in the far infrared  (450 and 850 \mm),
so the information about the infrared (IR) emission is rather complete.

The main results obtained from the interpretation of the line spectra (Blair et al)
are, in particular, 
that the emitting gas has high densities ($\geq$ 1000 \cm3), that the N/H relative abundance 
is $\sim$ 3.5 times higher than solar, and \Ha ~line profiles display both broad and narrow components.
A distance of  5 kpc  was consistently determined.

The   origin of Kepler SNR emitting features  is explained by
the interaction of the high velocity ejecta with the ambient
medium (Decourchelle \& Pretre 1999) and is very similar to that  
found for other SNRs , e.g. Cassiopeia A (Chevalier \& Oishi 2003), namely,
"that the observed shock wave positions and expansions can be interpreted in a model of supernova
interaction with a freely expanding stellar wind."
The  collision  with ambient matter leads to the formation of two shock fronts : 
a shock propagating outwards throughout the circumstellar/interstellar medium  and one
propagating  in reverse throughout the ejecta in the direction opposite 
to the propagation of the blast wave. The gas in
 the knotty regions  of Kepler SNR is ionized and heated to relatively high 
temperatures ($\leq$ 3 10$^7$ K)  downstream of the shock which
 expands  with velocities  of 1000 -1500 \kms.   
 The reverse  shock, on the other hand, 
 propagating with velocities of $\sim$ 50 \kms 
in the opposite direction, 
leads to lower gas temperatures.
The velocities are
reduced by the  density gradient of the ejected matter.
The circumstellar origin of the optical emitting gas  suggested that Kepler's
SN was actually of Type Ib (Bandiera 1987).

Very recently Morgan et al (2003) addressed  the problem of 
dust in Kepler SNR in terms of  its total mass and its origin from a massive star.
They claim that dust formation in SNe   is an important process,  but dust
in SNRs has been detected in small quantities.

In this work we will use the method adopted by e.g. Contini \& Contini (2003)
to calculate the dust-to-gas ratios in the starburst regions of luminous
infrared galaxies  and by Contini, Viegas \& Prieto (2004) for Seyfert galaxies.
Namely,
dust-to-gas ratios  are  calculated by  modeling 
the spectral energy distribution (SED) of the continuum in  the optical-IR 
frequency range.

At the shock front edge of the emitting nebulae, the gas is collisionally heated
to relatively high temperatures which depend on the shock velocity. 
Dust and gas  are coupled crossing the shock front,
mutually heating and cooling by collisions
(Dwek \& Arendt 1992, Viegas \& Contini 1994, Contini et al. 2004).
The temperature of the grains  depends  on the  temperature of the gas and,
therefore, the maximum temperature of dust depends on the shock velocity (Contini et al. 2004).
The observational data constrain the model which better  explains both
the bremsstrahlung and reradiation by dust. Particularly,
the ratio between  the dust radiation flux  in the IR and bremsstrahlung
 depends  on the dust-to-gas ratio.  

However,  the modeling of the continuum SED must be  cross-checked by the 
modeling of the line spectra, which is strongly constraining.

 In this paper I would like to  find out the physical conditions in 
the different  regions of Kepler SNR, as well as  the dust-to-gas ratios,
by consistent model calculations,   which take into account the results of previous investigations.

The SUMA code (Viegas \& Contini 1994, Contini \& Viegas 2001 and references therein), which is
adapted  to the calculation of gas and dust       
 spectra emitted downstream of a shock front, is  adopted.

The  general models are presented in Sect. 2. The models specifically calculated
for the Kepler SNR appear  in Sect. 3.1. The comparison of  the spectra with the data  is discussed
in Sect. 3.2  and cross checked by the fit of the continuum SED in Sect. 3.3.
Discussion and concluding remarks  follow in Sect. 4.

\section{The models}

\begin{table*}
\centering
\caption{The models selected from the grid}
\begin{tabular}{lrrrrrrrrrrrrrrr} \\ \hline
\ model & m1 & m2 &m3  & m4 & m5 &m6 & m7 &m8&m9   & m10\\
\hline
\ \Vs (\kms) &50 & 50 & 50 & 60 & 300&400& 700&  800& 1000&1000\\
\ \n0 (\cm3) & 3(3)&6(3)&9(3)&9(3)&300&300&60 &40  & 10&100\\
\ d/g (10$^{-14}$)& 1&  1&  1&   1& 1&  1& 4& 5 & 9&0.5\\
\ [OII]3727+3729 &  22.0 & 13.0 &  8.6 &  8.0 &  5.0 &  3.50 & 7.6& 12.7& 11.2&4.7 \\ 
\ [NeIII]3869+3968 &    .015 &   .010 &   .007 &   .33 &  1.58 &  1.890 &2.1 & 2.19 & 2.65&1.08\\
\ [SII]4069+4076 &   1.80 &  1.90 &  2.10 &  1.70 &   .90&   1.400 & 7.0& 0.8 & 1.5& 1.5\\
\ [OIII]4363 &    .006 &   .004 &   .002 &   .20 &  1.00&   1.110 & 1.04& 1.11 & 0.27&0.17\\
\ HeII 4684 &    .00 &   .00 &   .00 &   .00 &   .08&    .140 & 0.27 & 0.24 & 0.16& 0.3\\
\ [OIII]4959+5007 &    .12 &   .07 &   .05 &  2.89 & 12.30&  12.800 & 13.1 & 13.7 & 3.53& 2.17\\
\ [FeII]5159 &    .25 &   .20 &   .19 &   .20 &   .23 &   .450 & 0.2 & 0.25& 2.2& 0.7\\
\ [NI]5200 &    .02 &   .01 &   .007 &   .005 &   .06&    .140 & 0.09 & 0.13& 2.7& 0.14\\
\ [FeIII]5270 &    .34 &   .35 &   .34 &   .28 &   .30 &   .110 & 0.089& 0.11& 0.035& 0.015\\
\ [NII]5755 &   .54 &   .62 &   .63 &   .63 &   .24 &    .260&0.177& 0.22 & 0.1& 0.07\\
\ He I 5876 &    .11 &   .14 &   .116 &   .18 &   .33 &   .330 & 0.21& 0.23& 0.6& 0.2\\
\ [OI]6300+6364 &   .64 &   .48 &   .46 &   .50 &  1.10 &  2.450 &0.88 & 1.1 & 14.& 3.45\\
\ [NII]6548+6583 &   8.00 &  7.40 &  6.27 &  5.12 &  2.90 &  2.860 &3.25 & 4.14& 6.26& 3.5\\
\ [SII]6717&    .43 &   .24 &   .18 &   .10 &   .37 &   .630 & 0.6 & 0.93 & 16.7& 0.7\\
\ [SII]6730&    .90 &   .55 &   .41 &   .24 &   .78 &  1.330 &1.16& 1.7& 14.2& 1.5\\
\ [OII]732+7330&  10.00 & 12.30 & 12.00 & 13.00 &  5.50 &  5.700 &3.88& 4.6 & 0.89& 0.8\\
\ [SIII]9069+9532&   1.90 &  2.00 &  1.54 &  2.16 &  1.14 &  1.120 &1.0 & 1.17& 0.29& 0.13\\
\ [CI]9823+9849&    .30 &   .17 &   .15 &   .50 &  1.580 &   .160 & 0.39 & 1.3 & 8.8& 0.52\\
\ H$\beta_{abs}^1$ &      1. &   2. &   3.7  &   4.   &   1.4  &   1.7  & 0.53 & 0.33& 0.38& 9.13\\
\hline
\end{tabular}

\flushleft

  $^1$  10$^{-3}$ \erg

\end{table*}

\begin{figure*}
\includegraphics[width=88mm]{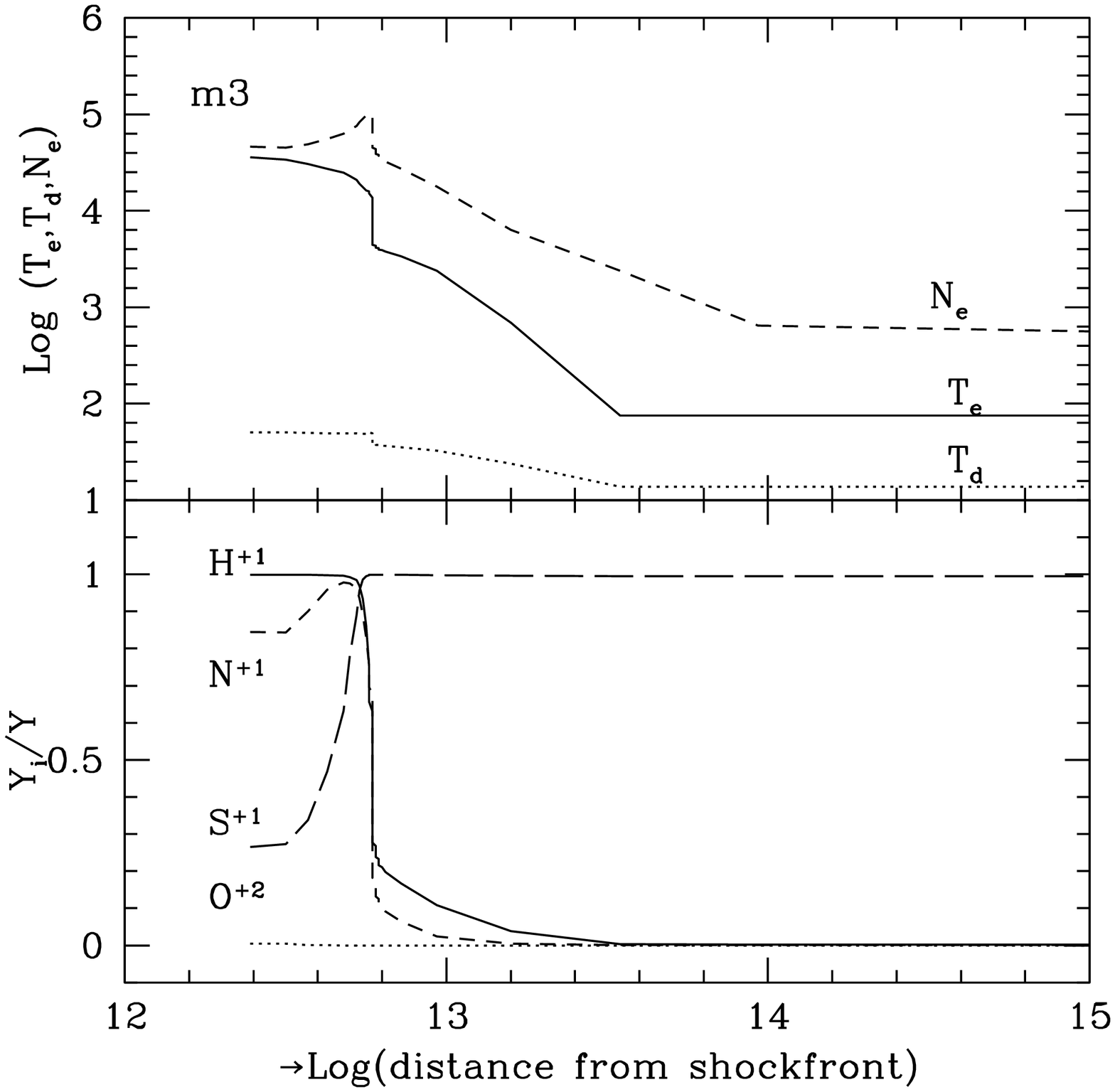}
\includegraphics[width=88mm]{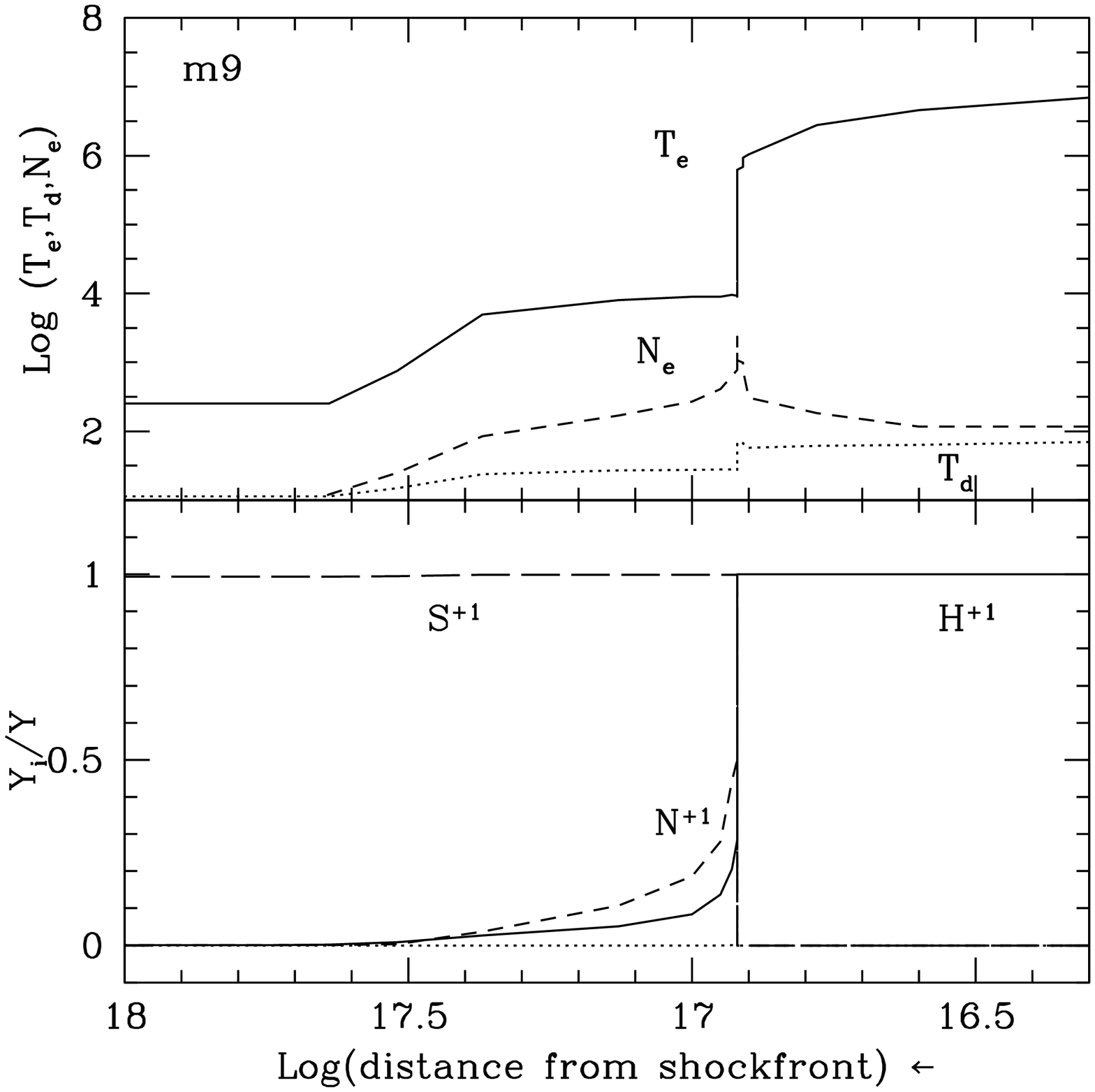}
\caption
  {The distribution of the electron temperature (solid line), of the dust temperature
(dotted line), and of the electron density (dashed line)
downstream (top panels) and of the fractional abundances of some significant ions downstream
(bottom panels) for model m3 (left diagram) and model m9 (right diagram).
H$^{+1}$/H : solid line; N$^{+1}$/N  : short-dashed line: S$^{+1}$/S : long-dashed line;
O$^{+2}$/O : dotted line }

\end{figure*}

The models   result from numerical simulations with
the code SUMA   that requires the following input parameters:
the shock velocity, \Vs, the pre- shock density, \n0,
the pre-shock magnetic field, \B0, 
the chemical composition of the gas, and a dust-to-gas ratio $d/g$ by number.
The initial  radius of the grains \agr is also required.
The geometrical thickness of the cloud, D, is 
an input parameter in the case of matter-bound models.
A plan-parallel symmetry is adopted with the calculations starting
at the shocked edge.
To perform the calculations a cloud is divided into a
number of slabs (up to 300) within which  the
physical conditions are  considered  homogeneous.
The geometrical thickness of
the slabs is automatically calculated by SUMA
according to the gradient of the temperature downstream, in order to calculate
as smoothly as possible the distribution of the physical conditions throughout the
cloud.  The first slabs in the
immediate post shock region show the maximum temperature which
depends on \Vs. These slabs can be relatively large because
recombination coefficients are lower the higher the temperature.
The geometrical thickness  of the slab closest to the shock front
is therefore determined by the model.

\subsection{Calculation of the spectra}

The  spectra emitted from  gas downstream  of the shock front,
are  calculated  with  a large number of equations,  
starting with those (e.g. Cox 1972)
 for the conservation of mass, momentum, and energy,   which lead to the
compression equation. Once  the physical conditions are determined in  a given slab,
 the ionization equations for  all the elements are resolved.
Then, the cooling rate is calculated   from the energy losses by free-fre, 
free-bound, and line emission, as well as by collisional
heating of dust grains  (Dwek \& Arendt 1992).

The  distributions  of the electron temperature and of the electron 
density  downstream  are thus  obtained,
as well as the stratification of the different ions for the different elements.

The structure of the gas downstream is determined by the shock. 
Near the shock front edge, the gas is  collisionally heated
to relatively high temperatures  (T $\propto$ \Vs$^2$) which determine
 the maximum of bremsstrahlung    at higher frequencies.
 Notice that  both free-free and free-bound  processes
are included in the calculation of the continuum. 

Due to  dust  collisional heating by gas, the maximum
temperature of dust depends on the shock velocity, which therefore
determines the frequency of the reradiation peak in the infrared.  An
analytical relation between the wavelength of the maximum intensity of
dust reradiation in the IR spectrum and shock velocity is given by
Draine (1981).

The grains are also heated by radiation. However, the only
radiation source in SNR is diffuse secondary radiation from the
hot slabs of gas, collisionally heated by the shock.
Diffuse radiation ionization rates prevail on collision rates only  when
the gas temperature has dropped to T$\leq$  10$^4$ K.
Radiation heating of grains, which is consistently calculated
by SUMA,  leads to low temperatures of the grains, $<$ 16 K
in high velocity clouds and  $<$ 30 K in low velocity clouds.

In models accounting for the  shock, sputtering
changes the initial grain-size distribution, creating
a deficiency of small grains compared to their pre-shock
abundances (Dwek, Foster, \& Vankura 1996).
The  grain size distribution depends on the shock velocity
and on the density of the medium.
The sputtering rate increases when the dust is in motion.
The present calculations account for the sputtering in the different zones
downstream of the shock front, so, the distribution of
the grain sizes along the cloud is automatically derived by SUMA
starting from an initial size. 

 Generally, three populations of grains are adopted to fit
the average interstellar extinction curve: large grains of silicate
with radius  \agr $\sim$ 0.01 -1.0 \mum, small graphite particles (\agr $<$ 0.05 \mum), 
and policyclic aromatic
hydrocarbons (PAH) which correspond to single molecules of 25 carbon atoms
and clusters of 10-20 molecules. Small graphite particles 
and PAH are underabundant with respect to the interstellar medium by 
factors of 10 or more  (Siebenmorgen et al 1999).
Moreover, they are rapidly destroyed by sputtering across the shock front
(Contini \& Contini 2003, Contini et al. 2004).
So, only silicate grains are considered.

Depending
on the temperature distribution in the cloud,  bremsstrahlung may
contribute to the infrared continuum
(see, for instance, Contini \& Viegas-Aldrovandi 1990, fig 2c).
 The intensity of dust IR emission
relative to bremsstrahlung depends on the  dust-to-gas  ratio.
 Usually the dust contribution dominates in the far infrared
range but in the near-IR, both processes may contribute to the
continuum (see  Sect. 3.3).

Finally, very seldom the
line and continuum spectra of the galaxies can be reproduced by
single-cloud models; on the other hand, multi-cloud models are
generally required to explain the multiwavelength spectrum
and are obtained as a weighted average of the single-cloud models. 
Multi-cloud models lead  to IR bumps wider than black body curves. 

The modeling of a spectrum consisting of only a few lines,  deserves
a further comment. In fact, it is   reasonable
only when   such a spectrum belongs to a series of spectra observed in 
neighbour  regions,  for a certain object. 
 In this case,  the  input parameters 
  which   are not likely to change in  neighbour regions, 
 e.g. relative abundances  and the pre-shock magnetic field,
are determined from  the   spectra rich in number of lines,
while, the shock velocity is roughly deduced from the line FWHM.
If the observations do not provide the lines whose ratio depends
on the  density of the emitting gas, the pre-shock density can be roughly
obtained from the modeling of  neighbour regions.
On the other hand,  the geometrical thickness of the cloud
is a free parameter because it depends on fragmentation.
Some times,  different combinations of the input parameters  lead to
very similar results for one  or two of the most significant line ratios.
So, the  final choice  of the model is determined by  consistency  
on a large scale. 
Nevertheless, spectra showing only  a few lines  should be  ignored, if they are not indispensable.

\subsection{Choice of the input parameters}

A first choice of \Vs ~comes from the FWHM of the observed line profiles
and, generally,  from  the intensity ratio  of  lines from different ionization levels. 
The range of \n0 ~is suggested by the [SII] 6717/6730 line ratios.
Notice, however, that the density of the line emitting gas  downstream  
is higher than \n0 by a factor $\geq$ 10 due to compression.

In matter-bound models the geometrical thickness of the nebula
is roughly determined by the intensity of low-level and neutral lines.

The magnetic field and the density are linked  through the compression equation
(Cox, 1972) which is calculated in each slab of
the gas downstream, leading to the density profile.
The range of \B0 ~is suggested by the results obtained for the other supernova
remnants (Cygnus Loop, Cassiopeia A, etc) and for the interstellar medium,
\B0=10$^{-6}$ - 10$^{-5}$ gauss
(Contini et al. 1980,  Contini \& Shaviv 1982, Contini 1987).
The choice of the pre-shock magnetic
field  in modeling is also  constrained   by the best   agreement
of  all the  calculated line ratios with the observational data. 

The models discussed here have been obtained assuming 
as a first choice  cosmic chemical abundances for  He, C, N, O, Ne,
Mg, Si, S, Ar, and Fe  (Allen 1973). 
The elements which reveal depletion from the gaseous phase
by trapping into grains are carbon, silicon,  magnesium, iron, etc.
Most of their lines are prominent in the UV and in the infrared spectra,
while the observed lines  are actually in the optical- near infrared.
Moreover, sulphur can be trapped into CS diatomic molecules.

 The dust-to-gas ratios are constrained by the SED of the continuum
in the IR range.
It is found that only  large grains (\agr $\sim$ 1 \mum) survive sputtering in high 
velocity shocks. 
The sizes of  grains adopted for low velocity shocks are 
constrained by the  data at long  wavelengths (Contini et al. 2004, fig. 7, left 
panel) and  \agr = 0.2 \mum ~ is obtained by comparison with SCUBA data.
 The contributions of dust emission from all the slabs downstream are summed up and lead 
to the IR bump.

\begin{table}
\centering
\caption{Comparison of calculated  with observed  line ratios (\Hb=1)
 from the whole remnant } 
\begin{tabular}{ l l l l l l l l l l l } \\ \hline
 line & obs.$^1$ &  obs.$^2$ & M1 & M2  \\ \hline
\ [OII] 3727+3729 & 6.74  & ($>$7)& 6.7 & 8.6 \\
\ [NeIII] 3869+3968 & - &  0.21&0.58& 0.58\\
\ [SII] 4069+4076 &2.87 & 1.37 &1.6 & 1.8  \\
\ [OIII] 4363 &0.37& 0.14 & .27 & .33 \\
\ HeII 4686 &- & 0.01 &0.06& 0.04  \\
\ [OIII] 4959+5007 &4.3&3.3 &3.6& 4.0 \\
\ [FeII] 5159 & - & 0.17 & 0.3  & 0.26 \\
\ [NI] 5199+5200 &-&0.04 & 0.1  & 0.07\\
\ [FeIII] 5270 & - & 0.04 & 0.2  & 0.29 \\
\ [NII] 5755&0.26 &0.11 & 0.4 & 0.5  \\
\ HeI 5876&0.15  & 0.12&0.18 & 0.19  \\
\ [OI] 6300+6364 &1.75& 1.52 &1.2 & 1.05   \\
\ [NII] 6583+6548 &7.16& 7.0&4.8 & 5.4   \\
\ [SII] 6717  & 0.5& 0.55 & 0.6  & 0.5  \\
\ [SII] 6730 &1.12& 1.12 &1.1  & 0.83   \\
\ [SIII] 9069+9532 &0.48& -& 1.3  & 1.4\\
\ [CI] 9823+9849 & 0.20 &-&0.3&0.46 \\
\ w(m1)       & - & - & 0 & 3  \\
\ w(m3)       & - & - &1.3&5  \\
\ w(m4)       & - & - &1 & 0 \\
\ w(m5)       & - & - & 0 &3  \\
\ w(m6)      & - & - &1 &3  \\
\ w(m7)       & - & - &0  & 1 \\
\ w(m8)       & - & - &1.3& 0 \\
\ w(m9)       & - & - & 0 & 1 \\
\hline
\end{tabular}

\flushleft

$^1$ Dennefeld (1982);  merged 1979 spectrum corrected for reddening  by c=1.69

$^2$ Leibowitz \& Danziger (1983):  dereddened by c=1.46

\end{table}

\section{Modeling  Kepler SNR spectra}

 The FWHM of the observed \Ha ~line
profile  suggests that  two kinds of shocked  clouds contribute to the spectra 
from  each observed region,  those with  high (\Vs $\sim$ 1000 \kms)   and those  with low velocities 
(\Vs $\sim$ 50 \kms).
The pre-shock densities are high enough as to obtain [SII]6713/[SII] 6730 $<$ 1.
Considering the adiabatic jump  at the shock front and that compression downstream increases with the shock
velocity, low velocity clouds  must be characterized by high pre-shock densities, while high velocity
clouds which propagate outwards  and have reached the outskirts of the SNR have lower densities.
This will lead to non radiative shocks characterized by very long recombination times of the gas downstream.
Recombination coefficients decrease at high temperatures and if
the densities are too low  to speed up the cooling process, 
the  temperatures within  the cloud will never  drop  below T= 10$^{5}$ K, which 
are critical  to the optical (and UV) line emission.  In other words, these models are matter-bound.
On the other hand,  X-rays which correspond to bremsstrahlung from high temperature gas  are 
actually observed,  revealing the presence of these clouds.

  Adopting  dust-to-gas
ratios as those predicted for Kepler's SNR by other  authors (e.g. Morgan 2003)
and confirmed  cross-checking the  results of  the present paper (Sect. 4), 
the energy losses by collisional
heating  of  grains prevail  downstream  of  high velocity
clouds, even characterized by a low  density,  leading to cooling times  much
shorter  than the age of the remnant (400 yr).
This is an important issue in view of the choice of the models.
In particular, d/g
should be  higher by at least a factor of 3  than normal values  (10$^{-14}$ by number)
and the grain  radius  large enough (\agr $\sim$ 1 \mum) to survive sputtering.

Due to fragmentation which follows from turbulence in shock dominated regimes, many different 
emitting nebulae (clouds) coexist in each region, so, to model the  corresponding observed spectrum, 
 different single-cloud models   must be
 summed up adopting relative weights in order to obtain the best
agreement with the  data. Notice that the structure of the observed regions is very complex,
and  a perfect fit  to the observed spectra by  some typical models is senseless.
So, in the following we will try to explain only the most significant line ratios for each of the observed spectra.

\subsection{The grid}

In Table 1 we present the single-cloud models selected from a large grid
 covering    all the possible  physical conditions which are consistent
with the Kepler SNR events.  Each of  the models  leads
 to the  rough fit of some of the  observed line ratios. We  reduced the number of  models to 
a minimum  of
prototypes, because modeling provides only an approximated picture.
The   ranges of the input parameters    were chosen 
 by  the  observational evidence and   
by a  first modeling of some significant line ratios
([OIII] 4959+5007, [OII] 3727+3729,
[OI] 6300+6364, [NII] 6548+6583, and [SII] 6717 and 6730), considering that,
finally, only   multi-cloud models  will be  compared with the data.

The  first three rows of Table 1 show the model input parameters. 
Then, the calculated line ratios
to \Hb=1 are given, followed by  the \Hb ~~absolute flux in the last row. 
 \B0=10$^{-5}$ gauss is adopted for all the models.
The models  in Table 1 are ordered   by increasing  shock velocities.

The selected models (Table 1) show three different types  characterized by
1)  low velocity shocks with \Vs $\sim$ 50 \kms and \n0 $\geq$ 3000 \cm3 (m1-m4),
2) intermediate shocks (\Vs=300,400 \kms) which propagate outwards and were decelerated by 
impinging on high density debris  of previously ejected matter (m5 and m6), and
3)  high velocity shocks with \Vs $\geq$ 700 \kms, \n0 $<$ 100 \cm3, and d/g $>$ 3. 10$^{-14}$ (m7-m9).   

The observed FWHMs  suggest that shocks with  \Vs $\sim$ 50 \kms must be accounted for (m1, m2, m3, m4). 
The corresponding pre-shock densities are $>$ 1000 \cm3 in order  to obtain
[SII] 6717/6730 $<$ 1  downstream. The  spectra 
show rather high [OII]/\Hb ~and very low [OIII]/\Hb. Interestingly, the [NII]6583+6548/\Hb
~line ratios are  similar to the observed ones, without adopting  N/H higher than cosmic,
while,  [SII] 6717/\Hb ~ and [SII] 6730/\Hb ~ are very low.
Notice that high {NII]/[SII] line ratios are  characteristic of the  observed spectra (Tables 2-6).
Actually,  the
first ionization potential of nitrogen   is higher 
 than  that of hydrogen, while the first ionization potential of sulphur is lower.
 So, the [NII\ and [SII] lines are emitted from gas  in different physical conditions, 
which  rapidly change  downstream within small distances (see Fig. 1, left diagram).

The  observed large FWHMs lead to models with \Vs $\geq$ 700 \kms and relative low \n0 = 30-60 \cm3
(m7, m8).
These models   show  [OIII]/\Hb, [OII] 3727+3729/\Hb, [NII]/\Hb, and [SII]/\Hb  ~ higher than  those 
calculated by models 
with \Vs  = 300 - 400 \kms and \n0  = 300 \cm3 (m5, m6), but show lower  [OII] 7320+7330/\Hb .

Model m9    shows peculiar line ratios, because a high \Vs ~leads
 to  a strong compression of the intermediate-level line (e.g. [OIII]) emitting  region,  
and to  a large  zone of  gas with \Te $<$ 10$^4$ K (Fig. 1, right)
which corresponds to  strong low level lines ([OI], [OII], [NII], [SII], etc).  
 Model m10 was added in Table 1
after discussing the SED of the continuum (Sect. 3.3).

The distribution of the electron temperature, of the electron density, and of the fractional abundance
of the most significant ions  as  function of distance from the shock front (in cm)  are presented in Fig. 1,
for models m3 and m9. Model m3 (left diagram) represents the nebula downstream of the reverse shock
which is characterized by  low \Vs ~and high \n0, while model m9 (right diagram) represents
the nebula downstream of the shock propagating outwards. Therefore, the shock front is shown
on opposite edges in the two diagrams. 
 
The emitting clouds are radiation bound.
The ionized region is rather narrow ($\sim$ 10$^{13}$ cm)
in  the low velocity model m3  (Fig.1, left panel).
On the other hand, model m9 shows a   high temperature zone of $\sim 10^{17}$ cm  (Fig. 1, right panel) which is
 reasonable compared to the  SNR diameter  (3.8 pc), calculated   adopting a Kepler SNR distance to Earth 
of 5 kpc (Blair et al.). 
The high  dust-to-gas ratio  (9 10$^{-14}$)  corresponding to model m9 leads to  a cooling time $<$ 60 yr
in  the downstream region.

\subsection{Line spectra}

\begin{figure}
\includegraphics[width=88mm]{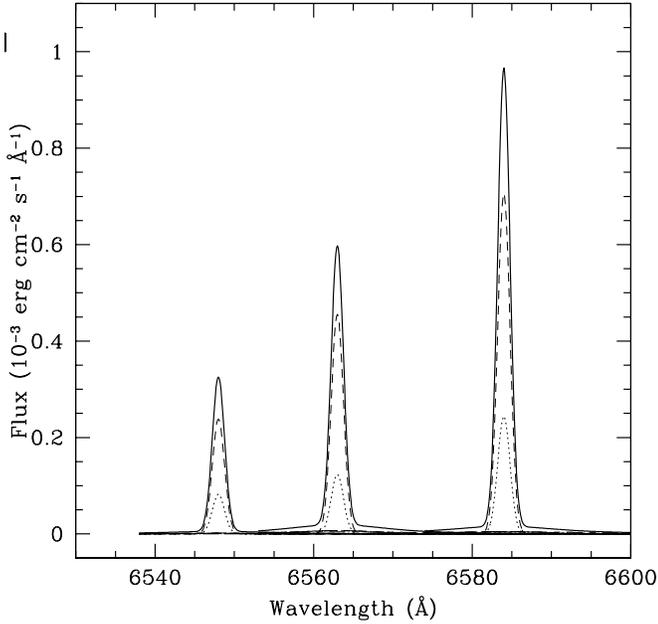}
\caption
{[NII] 6548, [NII] 6583, and \Ha line profiles (solid lines) calculated by model MB3
for region D3 observed by Blair et al (1991).
Dotted line correspond to model m1,  short-dashed line to m3.
The fluxes are  calculated at the emitting nebula.
}

\end{figure}

The data used  for modeling  the line spectra come from the following observations.
Spectrophotometry data of the bright-western knots covering the
spectral range 3700-10,500 \AA ~were obtained during 1979-1980 with the Boller
and Chivens spectrograph attached to the ESO 3.6 telescope and were
presented by Dennefeld (1982).
Observations in the optical range for single bright knots
in different positions and the integrated spectrum obtained in the
Anglo-Australian  3.9m Telescope come from Leibowitz \& Danziger (1983).
Blair et al. (1991) presented optical CCD/interference filter imaginery and long-slit CCD
spectrophotometry
of the Kepler SN  remnant obtained by the 2.5m DuPont telescope at Las Campanas Observatory.
Although  Blair et al.  data cover a relatively reduced wavelength range, the single spectra
observed in more than 20 regions, provide the tools for modeling the SNR extended area.

The mean reddening E(B-V) to Kepler's SNR is 0.9 $\pm$ 0.1 (Blair et al.), lower than 
E(B-V)=1.1 given by Leibowitz \& Danziger.

The comparison between calculated and observed (reddening corrected)  line ratios
to \Hb=1 is presented in Tables 2-6.   The spectra were 
observed by Dennefeld (1982), Leibowitz \& Danziger (1983), and  Blair et al. (1991).
 When the \Hb ~line is not observed we adopt \Ha/\Hb=3. 
The  weights of the corresponding single-cloud models
and the conversion factors (see Sect. 3.3) appear in the bottom of the tables.

 We will aim at fits such that the discrepancies between the calculated
line ratios and the observed ones are less than a factor of 2,
which accounts for observational errors and modeling approximation 
(atomic coefficients, etc.). 
However, not always the  agreement is satisfying for all the lines.
To have a better fit we should consider a  larger number of
single-cloud models (see, for instance,  region D55 in Table 5). 
Recall that  we are interested mainly in the calculation of the dust-to-gas ratios, therefore, 
in some regions which  are less constrained by  spectra poor in number of lines, 
the fit  is not sufficient, 
leading to   discrepancies beyond a factor of 2.

The two observed spectra  shown in Table 2 which  correspond to integrated data  by 
Dennefeld (1982) and Leibowitz \& Danziger (1983) are similar enough.
 The two multi-cloud models which better explain them correspond to  
 different  weights of the single-cloud models.
In Table 3 the models are compared  with observations in the different regions  
in the western side of the remnant by Leibowitz \& Danziger
(1983, Plate 3).

The  spectra observed by Blair et al. in the  different regions are generally satisfactorily explained by 
modeling (Tables 4-6). 
In the top of the tables  the  observed emission regions and their rough location
(D'Odorico et al. 1986) in the SNR are indicated.
 
The data  corresponding to regions D41,45 (Blair et al.) are not considered, because 
the spectrum is very poor in number of lines. Actually, it is the only position showing
very low densities ([SII] 6717/6730 $>$ 1) in  the emitting gas.  Blair et al 
claim that "this emission is not simply a superposition of a non radiative
and typical (for Kepler SNR) radiative filament."

The detection of [NII] 6583+ with a FWHM of $\sim$ 40 \kms (Sollerman et al, Table 1) 
from a Balmer dominated  region (e.g.  D49, D50) is explained by 
the reverse shock,
which  should  be summed up  adopting a very small weight compared with that of the high velocity model 
($>$ 1000 \kms). 
Sollerman et al  claim that the [NII] emission from D49 and D50 cannot be produced in a cooling
zone in Balmer - dominated knots. So, the remaining possibility is 
ahead of the Balmer-dominated shock, due to Alfven wave activity in a cosmic ray
precursor.
Table 1 shows that models calculated with low \Vs ~show very high [NII]/\Hb ~line ratios relative to [SII]/\Hb
, therefore, the only narrow lines observable are those of   [NII], and perhaps [OII], which
unfortunately, were not included in the spectra.
 Blair et al notice that
D47 and D48  with peak at 6584 ([NII]),  represent a transition filament implying a minimum shock velocity of 815 \kms.

Two spectra, D56 and P2D1 (P2,diff1) in Table 5, are explained by  single-cloud models, m5 and m6, respectively. 
 However, the modeling is less constrained by the small number of observed lines.

Summarizing,  the present modeling confirms that  N/H relative abundance higher than solar 
should be adopted,
in agreement with previous modeling ,  however,   by 
a factor  $<$2, which is lower than    the factor  of $\sim$ 3.5,  suggested by Blair et al.
This leads to  N/H $<$ 1.8 10$^{-4}$. Notice that models in group 1), corresponding to the slow shock,
and model m9
show rather high [NII] 6583+/\Hb ~ratios ($>$ 5). So, in the multi-cloud models which better fit the spectra, 
the ratios of  calculated to observed  [NII]/\Hb ~are closer to 1, the higher the relative weight of 
these models.
On the contrary, Fe lines are overpredicted by the models, indicating that Fe could be depleted 
from the gaseous phase. 

Tables 2 and 3 show that models with intermediate velocities (group 2)) have generally low relative weights.
They give  some contribution to the Leibowitz \& Danziger spectra (Table 3), and  even less to Blair et al  (Tables 4-6).
The  weights of the high velocity clouds (group 3)) in the Leibowitz \& Danziger and Blair et al. 
spectra   are low relative to 
those of the low velocity clouds (group 1) particularly for D9, D38,40, D63,64, SWD9, D9,10, and D25.

High velocity models   do not contribute to the spectra observed in some of the NE regions (D55, D56, D61, P2D1).
 Particularly,  radiative high velocity shocks are not found in these regions.
This indicates that the conditions in the outer region of Kepler SNR  are not homogeneous,  
showing different densities,  dust-to-gas ratios, and different grain sizes. 

On the other hand, high \Vs ~clouds are present in regions D3, D18, and D27.
The line profiles of  [NII] 6548, [NII] 6583, 
and of \Ha ~calculated by model MB3 for
region D3 (Table 4) are presented in Fig. 2. The line profiles of [NII] and \Ha ~(Fig. 2,)
show relative  fluxes in agreement with Blair et al. (1991, fig. 3). However, the calculated FWHMs
are  smaller, because the contribution of the high velocity gas
which leads to the best agreement of the  calculated  with the observed line spectra, 
appears as a socket in the  profiles. 
In fact, the absolute fluxes of the  \Hb ~lines (H$\beta_{abs}$  Table1, last row) and of the [NII] lines
([NII]/\Hb ~ $\times$ H$\beta_{abs}$)  are  low  for high \Vs ~models. 

\begin{table*}
\centering
\caption{Comparison of calculated  with observed line ratios (\Hb=1)}
\begin{tabular}{ l l l l l l l l l llll l } \\ \hline
 line &1 &ML1& 2&ML2 & 3&ML3& 4&ML4&5 &ML5&6&ML6\\
      &obs.$^1$ & calc. & obs. & calc & obs. & calc. &obs.&calc.&obs.&calc&obs.&calc.\\
\hline
\ [OII] 3727+3729 & (7) & 8.5& -    & -  & 11.    &10.5  &-    & -  & 7.   &9.5    &-     &- \\
\ [NeIII] 3869+3968 & 0.94& 0.6 & -   & -  & -     & -   &-    &  - & -    & -     &-     &- \\
\ [SII] 4069+4076 & 1.59 &1.87&0.97 & 1.9&   1.42 &1.87 &1.47 &  1.58&1.33 & 1.8  &-     & -  \\
\ [OIII] 4363 & 0.2  &0.3 &-    &-   &  0.27 &0.24  &0.12 &0.27& -    & -     & -    &-  \\
\ HeII 4686 & (0.01) &0.04&-    &-   & -     &  -   & 0.06&0.07& -    &  -    & -    & - \\
\ [OIII] 5007+ & 3.32&3.99&3.08 &3.2   & 2.7   &2.88  & 3.4 &3.59& 3.78 & 4.2  & 2.02 &2.26  \\
\ [FeII] 5159  & 0.25& 0.27&(0.03)&0.25  & 0.17  &0.28  & 0.29&0.3 & 0.12 & 0.33  & 0.05 & 0.27\\
\ [NI] 5199+ 5200 & 0.03 &0.07 &(0.07)&0.06 & 0.11  &0.10  & 0.05&0.1  &0.05   &0.117 & -   &- \\
\ [FeIII] 5270& 0.10& 0.29& - & -    & 0.04  &0.3   & 0.07&0.2 &0.05   &0.26&-   & - \\
\ [NII] 5755& 0.15    &0.49& - &  -  &0.20   &0.5   & 0.18&0.4 &0.09   &0.4& 0.05&0.5   \\
\ HeI 5876&   0.16    &0.18&- & -  &0.13    &0.17  &0.11 &0.18&0.06   &0.2& 0.08& 0.156\\
\ [OI] 6300+6364 & 1.5    &1.1& 1.03 &0.94&1.84    &1.11 &1.83 &1.21&1.41   &1.46 &0.84&0.98 \\
\ [NII] 6548+6583 & 6.94  &5.4& 6.13 &5.56&7.69    &5.9   &7.0  &4.82&8.2    &5.5& 5.97&6.1   \\
\ [SII] 6717  & 0.52  &0.50 & 0.36 &0.4 &0.72    &0.75  &0.51 & 0.61&0.66   &0.77& 0.67 & 0.64  \\
\ [SII] 6730 &  1.02  &0.83& 0.80 &0.75& 1.41   &1.00 &1.04 &1.08&1.42   &1.13 &1.28  &1.0 \\
\    c$^2$        &  1.51  &-  & 2.45&  -&  1.26  &-&   1.28& -&     1.35& - & 1.42&-\\
\   w(m1)     &   -    &3  &  -   &3  & -     &10   &  -  &  0 & -     &3     &-   & 15 \\
\   w(m3)      &  -    &5  &  -   &7   &  -     &9  &  -  & 2   & -     &2    & -   & 12 \\
\   w(m4)      &  -    &0  &  -   &0   &  -     & 0 &  -  & 1.5 & -     & 0   & -   & 0  \\
\   w(m5)     &   -    &3  &  -   &3   &  -     & 3   &   -&  0  & -     & 0  &  -   & 3  \\
\   w(m6)     &   -    &3  &  -   &3   & -     & 3   &  - &  1.5& -     &3   &  -   & 3   \\      
\   w(m7)     &   -    &1  &  -   &1 &  -     & 1   &  - &  0  & -     & 0  &  -   & 1 \\
\   w(m8)     &   -    &0   &  -   &0 &  -     & 7   &  - &   0 & -     & 0  &-   & 7  \\
\   w(m9)     &   -    &1  &  -   &1 &  -     & 4   &  - & 2   & -     &1    &-   & 4  \\
\hline
\end{tabular}

\flushleft

$^1$ observations by Leibowitz \& Danziger (1983, Table 3);

 $^2$  adopted for the dereddened spectra.

\end{table*}

\begin{table*}
\centering          
\caption{Comparison of calculated  with observed line ratios (\Hb=1)}
\begin{tabular}{ l l l l l l l l l llll l } \\ \hline
 line &   D3 &MB3& D9 &MB9  & D18 &MB18  & D34,35 &MB34 & D38,40&MB38   \\
 region & SW&-& SW&-&SW&-&SW&-&SW&-\\
      &obs.$^1$ & calc. & obs. & calc & obs. & calc. &obs.&calc.&obs.&calc.\\
\hline
\ [OIII]4959+5007 & 2.92&2.91 & 1.82& 1.84&  1.186 & 1.13 & - &-&   2.17& 2.15    \\
\ [NII] 5755&  0.22 & 0.5   & 0.18& 0.5 &   0.188 &0.5 &   - &-&   0.148& 0.57  \\ 
\ HeI 5876&  0.19 & 0.166 & 0.116& 0.15&   0.215 &0.17  & 0.15 :&0.196& 0.08& 0.15   \\
\ [OI] 6300+6364 & 0.79& 1.13 & 1.0&1.08&   2.58& 2.0   & 1.39 &1.16  & 0.734 & 0.64 \\
\ [NII] 6548+6583 & 5.87& 5.9 &   6.09&6.3&      9.2 &6.4  & 5.07&5.2 & 3.76& 6.11  \\
\ [SII] 6717  & 0.96 & 0.83& 0.67& 0.78&     1.28&1.9  &  0.39 &0.47& 0.2& 0.26    \\
\ [SII] 6730 &  1.59& 1.1  & 1.30 &1.1&   2.55&2.0  &  0.78 &0.85 & 0.38& 0.59  \\
\ H$\alpha^2$  & 69.&- & 164.&- & 74.2&- & 39.5&-  & 136.6& -\\
\ E(B-V)$^3$   &0.8969&-& 0.8034&-& 0.9648 &-& 0.9005$^4$ &-& 0.9077 &-\\
\ w(m1)       &- &12  & - &20  &-  &10 &- &3  &- &25 \\
\ w(m3)       &- &12  &-  &12  &-  &10 &- &7  &- &5 \\
\ w(m4)      &-  & 0  & - & 0  &-  &0  &- & 0 &- &8 \\    
\ w(m5)       &- &3   & - &3   &-  &0  &- &3  &- &1   \\
\ w(m6)      &-  &3   & - &3   &-  &1  & -&7  & -&2 \\
\ w(m7)      &-  &1   & - & 1  &-  &0  &- &1  &- &0\\     
\ w(m8)     &-   &1.8 & - & 1  &-  &4  &- & 0 &- &4 \\    
\ w(m9)      &-  &6   & - &6   &-  &15 &- &1  &- &0  \\
\ CF (10$^{-10}$) &-&1.5 &-&3. &-&0.5&-&5.9 &-& 2.3&\\
\hline

\end{tabular}

\flushleft

$^1$ observations by Blair et al (1991, Table 2)

$^2$ in 10$^{-15}$ \erg

 $^3$ reddening correction assumes intrinsic \Ha/\Hb=3.0

$^4$ a E(B-V)=0.9005  means that  exctinction of bright filaments is assumed.

\end{table*}

\begin{table*}
\centering         
\caption{Comparison of calculated  with observed line ratios (\Hb=1)}
\begin{tabular}{ l l l l l l l l l llll l ll} \\ \hline
 line &  D63,64&MB63 &  D55&MB55$^1$& D56 &MB56&D61 &MB6& P2D1  &MBD1&  SWD9&MBD9\\
 region&SE &-& NE&-& NE&-&NE&-&NE&-&SW&- SW&-\\
      &obs.$^2$ & calc. & obs. & calc &obs.&calc.&obs.&calc&obs.&calc.&obs.&calc.\\
\hline
\ [OIII] 4959+5007 & - &-& 0.17+ &0.79&- &-& 1.15& 1.74& - & - &-&-\\
\ [NII] 5755&  - &-   & - & - & - &-&   - & -&-&-&-&-\\
\ HeI 5876&  -  & -  & 0.037 & - & -&- & - & - & - &-&-&-       \\
\ [OI] 6300+6364 & 2.97:&2.06& 0.36+:&0.5& -&-&  0.53 :&0.72 & 3.64 & 2.45 & 1.24&1.1 \\
\ [NII] 6548+6583 & 6.19&4.03&   1.97 &3.9  & 4.08 &3.   & 5.8 &5.8 &   0.8 & 2.86&  9.88&4.14\\
\ [SII] 6717  & 0.56&0.61 &  0.092&0.06  & 0.38 &0.37  & 0.24&0.24&-& -& 1.35&0.93\\
\ [SII] 6730 &  1.28& 1.25& 0.20 &0.14 & 0.91  &0.8&  0.64 &0.53&-& - &  2.68&1.7\\
\ H$\alpha^3$  & 12.6  &-  &33.4&- & 22.9 &-& 18.2 &-& 20.6 &-& 34.8 &-\\
\ E(B-V)$^4$   & 0.9005 &-& 0.8926&-& 0.9005&-& 0.9005&-& 0.9005&-&0.9005&-\\
\ w(m1)       &- &  5   & - &0 &- &0 &- &0 &- &0 &- &4  \\
\ w(m3)       &-&  0     &- &0 &- &0 &- &3  &- &0 &- &0\\
\ w(m5)       &-&  0     &  -&0 &- &1   &- &0 &- &0 &- &0\\
\ w(m6)       &-& 1      &- &0 &- &0 &- &1 &- &1  &- &0\\
\ w(m7)      &-&   0     &- &0 &- &0 &-&0  &- &0 &- &0 \\
\ w(m8)      &-&   0     &-& 0 &-& 0 &-&0  &-& 0 &- &1 \\
\ w(m9 )      &-&0.01     &- &0 &- &0 &- &0 &- &0 &- &0   \\
\hline
\end{tabular}

\flushleft

$^1$ \Vs=55 \kms and \n0=2 10$^4$ \cm3

$^2$ observations by Blair et al (1991, Table 2)

$^3$ 10$^{-15}$ \erg

 $^4$ see comment about E(B-V) in the bottom of Table 4.

\end{table*}

\begin{table*}
\centering
\caption{Comparison of calculated  with observed line ratios (\Hb=1)}
\begin{tabular}{ l l l l l l l l l llll l l} \\ \hline
 line       &   D9,10 &MB10  & D25&MB25 &D27&MB27& D3&MB3* & D9 &MB9* & D18 &MB18* \\
 region     & SW &-&SW &-&SW &-& SW &-& SW &-\\
      &obs.$^1$ & calc. & obs. & calc & obs. & calc. &obs.&calc.&obs.&calc&obs.&calc.\\
\hline
\ [OIII] 4959+5007 & 0.5+; &0.4  & 0.65+&0.87 & 3.18&3.2  & 2.91 &3.0& 1.84&1.4  & 0.51+&1.13\\
\ [NII] 5755&  -       &-    & 0.12&0.5  & 0.20&0.4   & 0.5 &0.2  & 0.5 &0.18 & 0.17&0.5   \\    
\ HeI 5876&  -         &-    & 0.12&0.139 & 0.15&0.15   & 0.17 &0.19 & 0.15&0.2 &  0.18&0.17\\ 
\ [OI] 6300+6364 & 1.71    &1.6  & 1.78&1.32   & 1.36&1.16  & 1.13 &1.2& 1.08&1.3 & 2.33&2.0   \\
\ [NII] 6548+6583 & 7.95   &7.88 &   7.55&7.75& 7.18&7.1  & 5.9  &3.2& 6.3 &3.9 & 8.77&6.4  \\
\ [SII] 6717  & 0.92   &1.58  & 0.815&1.2  & 0.57&1.0  & 0.83&0.85  & 0.78&0.7&  1.32&1.9  \\
\ [SII] 6730 &  1.97  &1.84  & 1.67&1.6   & 1.18&1.5  & 1.1  &1.4& 1.32&1.1  & 2.55&2.0\\
\ H$\alpha^2$ & 213.4 &-      & 127.&-    &1026.7&- & 83.&-  & 215.8&- & 56.&- \\
\ E(B-V)$^3$      & 0.9005 &-& 1.0277 &-& 0.8097 &-& 0.8804 &-& 0.8048&-& 0.9186&-\\
\ w(m1)        &  -    &5     &  - & 7   &- &9    & -    &12 & - &20   & -  &10   \\
\ w(m3)        &  -    &0     &   -& 0   &- & 0  & -    &12 &-  &12  & -& 10 \\
\ w(m5)        &  -    &0     &   -& 0   &- & 0  & -    &3   &-  & 3   & -& 0 \\
\ w(m6)        &  -    &0     &   -& 0   &- & 0  & -    &3  &-   & 3  & - & 1  \\
\ w(m7)       &  -    &0      &   -& 0   &- & 0  & -    &1   & - & 1  & - & 0 \\
\ w(m8)       &  -    &0      &   -& 1   &- &8   & -    &18  & - & 1  & - & 4  \\
\ w(m9)       &  -    &1      &   -& 1   &- &1   & -    &6   & - & 6  & - &15  \\
\ CF (10$^{-11}$)&-& 7.&-& 4.5&-& 46.&-&-&-&-&-&-\\ 
\hline

\end{tabular}

\flushleft

$^1$ observations by Blair et al (1991, Table 2)

$^2$ 10$^{-15}$ \erg

 $^3$ see comment about E(B-V) in the bottom of Table 4.

\end{table*}

\subsection{The SED of the continuum}

\begin{figure}
\includegraphics[width=88mm]{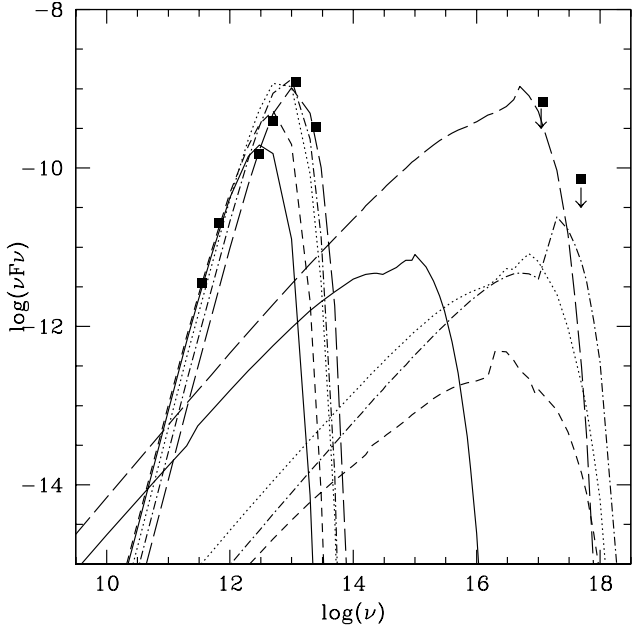}
\caption
  {
Comparison of single-cloud models with observational data
(black squares).
m3 : solid lines; m6 : long-dashed lines; m7 : dotted lines;
m9 : short-dashed lines; m10 : dash-dotted lines}

\end{figure} 

 The data in the X-ray  come from Becker et al. (1980), Tuhoy et al. (1979), 
 in the radio from  DeLaney et al (2002) and in the IR  from Arendt (1989),
Saken et al. (1992), and Braun (1997).
Very recently the SCUBA data at 450 and 800 \mm ~by JCM telescope on Mauna Kea were presented by Morgan et al. (2003).

Indeed the data sets in the radio and IR ranges  are quite complete and
allow  the modeling of gas and dust emissions (Figs. 3 and 4). However, data in the optical-UV range
are lacking. As a first check, we will constrain the models by the X-ray data, although they are
upper limit.

The modeling of the continuum  SED  must be consistent with that of the line spectra.
The SED  shows different contributions to the flux in the different
frequency ranges.
The gas  which is heated in the immediate post shock region to  
relatively high temperatures, 
cools down downstream   with cooling-rates which depend strongly on the gas density, on the relative
abundances of the most  coolant elements, and on the dust-to-gas ratio.
So, the bremsstrahlung emitted from  gas in different conditions covers a large range of frequencies,
from radio to X-ray.
Diffuse radiation from the hot gas maintains the gas ionized and heated to T $\leq$ 10$^4$ K
in a relatively large zone downstream.

The optical-UV radiation flux  is emitted from gas at T = 10$^4$-10$^5$ K, while the soft X-ray
domain  shows the bremsstrahlung  from high temperature gas. 

The IR bump  corresponds to reradiation by dust.
Mutual heating  of dust and gas throughout the shock front and downstream
leads to dust temperatures which depend on the gas temperatures and, thus, on \Vs.
Therefore, the frequency corresponding to the peak of dust reradiation  is generally  determined 
by \Vs ~(see Contini, Viegas \& Prieto 2004). 
The intensity of the IR flux depends, however, mainly on d/g.
The IR bump defined by the observations must be disentangled into the different  components 
corresponding to the different single-cloud models  which  are used to explain  the line spectra.

The modeling procedure starts by  fitting  the bremsstrahlung
which indicates the range of the shock velocity, consistently with
 the frequency peak in the IR.
Then, the IR bump is shifted along the Y-axis  
by changing the dust-to-gas ratio (Contini et al. 2004),
in order to obtain the best fit to the data.

In the radio range,  synchrotron radiation generated  by
Fermi mechanism at the shock front, generally dominates in SNRs.
Recall that the character of optical and X-ray magnetobremsstrahlung resembles that of
the radio counterpart, but to produce radiation  at optical and X-ray frequencies,
the  electrons must possess considerably higher energies than for  radiation
in the radio range, or the
strength of the magnetic field must be higher
(Ginzburg \& Syrovatskii 1965).
So,  the synchrotron radiation  is cut at 10$^{13}$ Hz  in Fig. 4 as an upper limit.

The SEDs of some significant models  are presented in Fig. 3. 
The  fluxes  are shifted in order to  fit  as much as possible the data.

Model m3  is  selected as  representative of
low velocity shocks (group 1), model m6 of intermediate shocks (group 2), model m7 of high velocity 
shocks (group 3), and 
model m9  is selected  because it is characterized by a high \Vs, a low \n0, and a high d/g. 
We have added in Fig. 3
the continuum SED of  model m10 corresponding to \Vs=1000 \kms, \n0=100 \cm3,
d/g=5 10$^{-15}$, \agr=1 \mum, and D=2 10$^{16}$ cm,  which is matter-bound. In fact, the gas
downstream contains only
 the   high temperature region, so,   the cooling time  is  smaller than
 2 yr. The corresponding radiation-bound model with D=10$^{18}$ cm (model m10 in Table 1) would lead
to cooling times of  $\sim$ 400 yr,  i.e.  of the same order of the remnant age,
because d/g is low. 
Model m10 shows that non radiative shocks can contribute strongly to the soft X-rays and
to IR radiation. However,  in region dominated by  grains small enough to be destroyed
by sputtering,  this type of models contributes only to the bremsstrahlung.
 
The  fluxes of models m7 and m9 displayed in Fig. 3  are   multiplied by a factor of 4. 10$^{-11}$ in order to fit the data.
Those of models m3 and m6 are multiplied by factors of 8.1 10$^{-9}$ and 4. 10$^{-9}$, respectively, and the
flux of model m10  by a factor of 1.6 10$^{-10}$.
Recall that models are calculated at the nebula while the data are observed at Earth.
 Therefore, conversion factors  which account  for the distances 
of the   observed regions from the  remnant center and of the  remnant to Earth 
must be considered. The modeling of the line spectra shows (Sect. 3.2) that the 
observed regions  are explained by multi-cloud models, so, 
 in this first analysis of  single-cloud models,  only the ratios of these factors are significant
because they show  the ratio of  the flux intensities   from  different models.

Fig. 3 shows that  the IR flux calculated by model m3   peaks at lower frequency  than  
model m6 because \Vs ~is lower.
The  high post shock temperature region corresponding to model m9 with \Vs= 1000 \kms,
is  reduced by the  strong cooling due to the  high d/g, so,   the peak in the IR, as well as the bremsstrahlung peak,
appear  shifted to  lower frequencies.
The bremsstrahlung of model m3  corresponding to low \Vs  ~is not constrained by the soft X-ray upper limits, 
therefore,
the dust-to-gas ratio cannot be confirmed. Recall that in both m3  and m6  
d/g = 10$^{-14}$ by number (corresponding to 4 10$^{-4}$ by mass for silicates) is adopted.
Fig. 3 shows that   the calculated IR bumps of   the models fit most of the  IR data well enough to confirm that these  
types of clouds contribute to  the SED of Kepler's SNR.

However,  the observational data correspond to the    whole SNR.
So, as a second step in modeling
the continuum SED, we compare the data with a model which accounts for  the
most   luminous regions observed by Blair et al,
 (D3, D9, D18, D34,35, D38,40, D9,10, D25, D27).  
Therefore, first,  the continuum corresponding to the multi-cloud  model which explains
the line spectrum is calculated for each  observed region. 
Then, the results  are summed up adopting   conversion factors   (CF) which
  are the same for the continuum  and   line spectra from each region.
The CF are, therefore, calculated  from the observed \Ha ~fluxes   given by 
Blair et al in the different regions, and  included in Tables 4-6, 
and from the  the calculated 
\Hb ~fluxes, which are given in Table 1 for the different models. \Ha/\Hb=3 is adopted.
The calculated CF for each region   appear in the bottom of Tables 4-6.
 Multiplying these  CF   by the   continua emitted from  each region and summing up,
the total continuum SED  is approximatively calculated for the Kepler SNR 
and can be  compared with the data.

In Fig. 4  the  bremsstrahlung is  indicated by a  solid line, while dust emission by a dashed line. 
We have separated
the dust emission curve from the bremsstrahlung to better understand their roles.
Model m10,  represented by dotted lines, has not been  included in the sum.

ISOCAM observations presented by Douvion et al (2001) in different regions, cover the range between 6.5 \mum ~ and
16 \mum. We have  summed up the fluxes from the different regions
 corresponding to  12 \mum ~and 16 \mum  ~and they are shown in Fig. 4 for comparison. 

To  further check  whether the calculated total bremsstrahlung in the optical range is consistent with 
the observations, 
 the continuum flux  at 5400 \AA  ~is roughly deduced from the  spectrum   observed  by 
Leibowitz \& Danziger (1983, Fig. 3)
and  is  indicated as  an open square in Fig. 4.
The constraint of the continuum SED is acceptable  and confirms that the  dust-to-gas
ratios adopted in single-cloud models,   are reliable.
Nevertheless, Fig. 4 shows that the two IR data  at 60 and  100  \mum  ~are overpredicted by the models.
A better fit could  be obtained  considering the observation error and/or
 by reducing the  weight  of models m7-m8  and/or  by reducing the
d/g ratio  of these models.  A lower d/g   leads to large cooling times downstream  of  models m7 and m8,
so that the model parameters should be rearranged destroying the good fit the line spectra.
A lower weight, too, leads to changes in some multi-cloud line spectra.
Indeed, the  fits of the observed continuum and of the  line spectra could be improved separately.
However,  model results for continuum and line spectra are cross-checked, leading to the  best modeling
of both.

The synchrotron radiation calculated in the radio range is  represented by  a thin solid line in Fig. 4.
DeLaney et al. (2002) find
a mean value in the spectral index of -0.71  for synchrotron radiation in the radio range,
which  generally originates from turbulence and magnetic field enhancement at the shock fronts
in SNR.  The  present modeling  shown in Fig. 4, adopting Fermi mechanism
at the shock front (Bell 1978) leads to a spectral index of 0.75, which  explains satisfactorily the data.
Moreover, the difference in structure between the flat and steep spectra of radio emission 
is explained by DeLaney et al. by a  partial decoupling of the forward and reverse shock.
Although we find that both types of shocks  act in nearly all the observed regions,
it is difficult to predict how much they have farthened at the time of observations.

Summarizing, it is found by modeling that dust-to-gas ratios  are 4 10$^{-4}$ by mass  
in clouds corresponding to low  velocity  shocks and even lower in intermediate velocity clouds.
On the other hand, d/g are relatively high  in clouds corresponding to  high velocity  shocks, 
by factors of 4, 6 , and   9 for \Vs   ~of 700, 800, and  1000 \kms,
respectively.

\begin{figure}
\includegraphics[width=88mm]{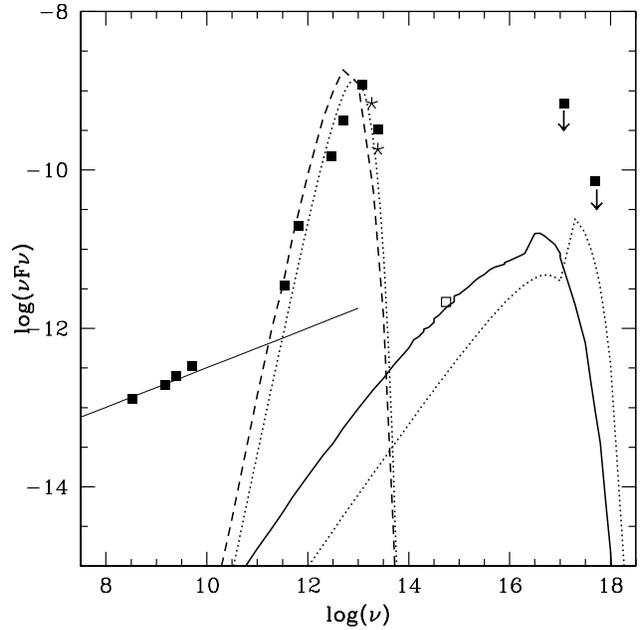}
\caption
    { 
Consistent modeling of the Kepler SNR continuum  SED
by the summed model (thick solid line : bremsstrahlung;
thin solid line : synchrotron radiation;
dashed line : IR emission) and model m10 (dotted lines).
The observational data are represented by black squares.
The datum at 5400 \AA ~from Leibowitz \& Danziger (1983) is represented
by an open square. The asterisks refer to Douvion et al (2001a)}

\end{figure}

\section{Discussion and concluding remarks}

In this work the calculation of dust emission  is carried out consistently with emission from
gas for  both  line and continuum spectra, leading to the evaluation of the dust-to-gas ratios
in   the Kepler SNR.
A successful application of this method (Contini et al 2004)  was presented
for luminous IR galaxies (Contini \& Contini 2003) and for AGN (e.g. Contini \& Viegas  2000).

The  two  component emission  observed in  the FWHM of the   \Ha ~line profiles
corresponding to velocities of $\sim$ 50 \kms and $\sim$ 1000 \kms is  confirmed by modeling.
They correspond to the reverse and the expanding shocks, respectively,   and are explained
by the supernova interaction with a stellar wind.

 A grid of single-cloud models is calculated. The models  cover the set  of physical conditions
in the emitting gas that are revealed  by the line ratios and by the FWHM of the line profiles. 
Multi-cloud models which account for the reverse and expanding shock in each region, 
are adopted in order to fit the line spectra in the different observed regions
and the corresponding continuum SED.
The total continuum  which results from summing up the contributions  of the most luminous
regions, is compared with the observed  continuum SED.
Dust emission  in the IR is consistently calculated.
The ratio between    dust emission  and   bremsstrahlung in the IR range
depends on the dust-to-gas ratios adopted in the models. 

The  high dust-to gas ratios  which are found  by modeling the IR  bump in  models corresponding 
to high velocity shocks, speed up
the cooling rate downstream, leading to radiative  shocks even   in  a 
relatively low density gas.
It is found that radiative high velocity shocks are absent in the NE region of the remnant,
suggesting that  d/g is low and/or the grains are small enough to be destroyed by sputtering.
This result may indicate that dust  is not homogeneously distributed  throughout the remnant.

Moreover, Kepler SNR  morphological structure in the different wavelengths 
can be   explained by model  results.
In fact, the X-ray emission has the same shell-like morphology and is qualitatively similar 
to the radio emission (Blair et al, DeLaney et al.), 
confirming that both radio  and X-ray are created at the shock front of the expanding
shock. Indeed, the   high temperatures ($>$ 10$^{7}$ K)  of the gas in the immediate
post-shock region downstream, correspond to the X-ray emission.

 The \Ha ~and IR images are  also similar (Blair et al.) and are similar to the
  X-ray image  as well.
Particularly,  the striking similarity between the \Ha ~and IR images  
suggests that the 12 \mm ~thermal
dust emission  and  the optical emission  have the same origin
(DeLaney et al. 2002)  confirming that    dust  grains  are collisionally heated by the gas 
 across the shock fronts and downstream.
 The  regions of dissimilarity  could be explained by the  
inhomogeneous distribution of  dust grains.

Notice that the
 modeling of the Kepler SNR through the line and continuum spectrum analysis
in the different regions is presented in previous sections  with the aim of  calculating the
dust-to-gas ratios. Actually, this work   has been inspired   by 
Morgan et al (2003) who claim that
dust formation in supernovae is required to be an 
important process relative to the age of the Universe. 

Morgan et al calculate that the maximum  dust  mass swept up by the SNR is $\sim$ 10$^{-3}$ \msol,
assuming a  maximum density  n = 0.1 \cm3 and a gas to dust ratio of 160.
However, adopting a  higher n (100 \cm3) the dust mass swept up is $\sim$ 1 \msol.
Douvion et al (2001) claim that by using  dust grains of 'astronomical silicates'
both mid-IR  ISOCAM and IRAS data can be fitted by a single grain temperature of 107 K,
while previous models required two dust components : a hot dust of 140 K and a
cold one of 54 K (Saken et al. 1992).
The total mass calculated by them is 10$^{-4}$ \msol.

The  model presented in this work  shows that  two  main  types of dust  lead to a good fit of
the IR bump consistently with the bremsstrahlung. 
Large  grains with \agr=1 \mum ~ with d/g $<$ 4 10$^{-3}$ survive sputtering downstream 
of the expanding shock and  are collisionally heated to a maximum
temperature of 125 K. Smaller grains (\agr=0.2\mum)  downstream of the reverse shock  
(Fig. 1) with   dust-to-gas ratios
of  $\sim$ 4 10$^{-4}$ are heated to a maximum temperature of $\sim$ 50 K.
 
Adopting a SNR  diameter of 3.8 pc, a pre-shock density $\sim$ 50 \cm3  by the average of models
m7, m8, m9 (Table 1) (Sect. 3.3), 
 we obtain a swept up gas mass of  40 $\it ff$ \msol and a maximum dust mass 
of   0.16 $\it ff$ \msol, where $\it ff \leq 1$ is the filling factor.

 Such  relatively high  pre-shock densities in single clouds are  justified  by
the relatively young age of Kepler SNR.
In older SNR  (e.g. the  Cygnus Loop), \n0 is less than 10 \cm3 (Contini et al 1980).
Moreover, the clumpy aspect of the remnant suggests that average densities on large scales
are not realistic and that $\it ff$ should be $\leq$ 0.1.

Comparing with  other SNR,
 Douvion et al (2001a) found that a dust mass of 10$^{-4}$ \msol  results from the warm component
(140 K as for Kepler SNR) and of 4 10$^{-3}$ \msol for the cold component (55 K) of Tycho.  
They claim that the Crab nebula
IR emission is dominated by synchrotron radiation and no dust is detected.
In Cassiopeia A the dust is  made  of more  components,  e.g. quartz and aluminum oxid, besides silicates
which have a mass of 7.8 10$^{-9}$ \msol (Douvion et al 2001b). 

\section*{Acknowledgments}

I am very grateful to  an anonymous referee whose precious comments helped to improve the paper
and to Elia Leibowitz for fruitful discussions.

\label{lastpage}

\end{document}